\documentclass[english]{sig-alternate}
\usepackage[T1]{fontenc}
\usepackage[latin9]{inputenc}
\usepackage[letterpaper]{geometry}
\usepackage{array}
\usepackage{rotfloat}
\usepackage{amsmath}
\usepackage{amssymb}

\makeatletter

\providecommand{\tabularnewline}{\\}

\usepackage{multirow}

\makeatother

\usepackage{babel}

\begin{document}
\numberofauthors{1}
\author{
\alignauthor Daniel Gayo-Avello\\
\affaddr{University of Oviedo}\\        
\affaddr{Despacho 57, planta baja, Edificio de Ciencias}\\
\affaddr{C/Calvo Sotelo s/n 33007 Oviedo (SPAIN)}\\        
\email{dani@uniovi.es}
}

\title{All liaisons are dangerous\\
when all your friends are known to us}
\maketitle
\begin{abstract}
Online Social Networks (OSNs) are used by millions of users worldwide.
Academically speaking, there is little doubt about the usefulness
of demographic studies conducted on OSNs and, hence, methods to label
unknown users from small labeled samples are very useful. However,
from the general public point of view, this can be a serious privacy
concern. Thus, both topics are tackled in this paper: First, a new
algorithm to perform user profiling in social networks is described,
and its performance is reported and discussed. Secondly, the experiments
--conducted on information usually considered sensitive-- reveal that
by just publicizing one's contacts privacy is at risk and, thus, measures
to minimize privacy leaks due to social graph data mining are outlined.
\end{abstract}



\section{Introduction and Motivation}

Graph Labeling is the task of assigning labels to the vertices or
edges of a graph. Because social networks are usually represented
as graphs, vertex and edge labeling algorithms can be applied to them
straightforwardly. In the former case the individuals in the network
are labeled, while in the later the labels are assigned to the relationships
between them. 

In that context, labeling algorithms can exploit a property of social
networks: the tendency of people to relate more likely with those
sharing similar traits, or \emph{homophily}. This phenomenon is pervasive
to very different social networks, and it has been revealed that a
number of personal characteristics --such as race and ethnicity, age,
religion, education, occupation, or sex-- induce homophilous relationships
\cite{mcpherson2001birds}.

Thereby, homophily can be used both to cluster similar individuals
within a network, or to infer attribute values for every individual
from his neighbors' characteristics. In the first case --community
detection-- it is not necessary to know anything about the members
of the social network except for their relations. In the second, the
attributes of interest are needed and, hence, part of the individuals
in the network must have known values for them --i.e. they must be
labeled. Thus, from a machine learning perspective, the first is an
unsupervised problem while the second is semi-supervised. 

It is well-known that online social networks (OSNs) %such as Facebook
ask their users for personal information %--e.g. birth date, sex, or marital status--
and that many of those users happily provide it. Hence, part of the
users in OSNs are labeled and semi-supervised approaches can be employed
to label the rest of the members of the network.

This paper describes a new semi-supervised algorithm to perform user
profiling in social networks. A number of experiments conducted on
Twitter\emph{ }data are reported, and the privacy implications discussed.
Certainly, there exist a number of semi-supervised methods to label
partially labeled social graphs. Hence, a later section reviews those
most highly related to the method here proposed, and points out the
main differences between this and them. Moreover, previous works regarding
the privacy implications of social graph mining are also discussed.
In that sense, this paper focuses on active measures the users can
adopt and, thus, it outlines a protocol to minimize leakages due to
graph data mining.

\section{The\emph{ }M\lowercase{c}C-Splat\emph{ }Algorithm}

\emph{McC-Splat}%
\footnote{Mnemonic for Multiclass Classification using Soft Labeling Propagation
and Automatic Thresholding.%
} is an iterative algorithm to perform vertex labeling on a partially
labeled social network. It is a multiclass classifier, that is, each
attribute can have more than two classes and, in fact, in addition
to the predefined attribute values --e.g. \emph{female} and \emph{male}
for \emph{sex}-- an extra class, \emph{unknown}, is also required
for each attribute. 

Needless to say, individuals have a number of different attributes
--e.g. sex, age, or marital status-- and, hence, each person would
have got a \emph{profile} comprising such attributes with their corresponding
values. The \emph{McC-Splat} algorithm can simultaneously propagate
the values for each attribute in the users' profiles but, for the
sake of clarity, the following description just covers the single
attribute case.

\emph{McC-Splat} works on a directed graph $G=\{V,E,C,A\}$ where
$V$ is the set of vertices --i.e. individuals, $E$ denotes the edges
--i.e. relationships between those individuals, $C=\{c_{1},\ldots,c_{m}\}$
are the different classes each individual's attribute can take, and
finally $A$ is the set of attribute weight vectors for the individuals
in $V$. 

By using attribute weight vectors it would be possible to model overlapping
classes; nevertheless, all of the experiments reported here were conducted
with disjoint classes. Moreover, the weights can be seen as a proxy
for the user's likelihood to belong to a given class --including the
\emph{unknown} one-- although weights are not probabilities in a strict
sense. The following formalization describes the $A$ set:

\noindent \begin{center}
$A\subset\left[0,1\right]^{m+1}|m=\left|C\right|,\forall\mathbf{a}\in A:\left|\mathbf{a}\right|=1$
\par\end{center}

Given that $G$ is partially labeled, $V$ is divided in two disjoint
sets: the set $V^{K}$ of known vertices --i.e. those individuals
with a known class value for the attribute-- and the set of unknown
vertices $V^{U}$ --i.e those belonging to the \emph{unknown} class.
Because the attribute values can be taken from $m$ different classes
this can be formalized as:

\noindent \begin{center}
$V=V^{K}+V^{U}$
\par\end{center}

\noindent \begin{center}
$V^{K}=\left\{ v_{i}\in V|\exists j:1\leqslant j\leqslant m,\mathbf{a}_{\mathbf{i}j}=1\right\} $
\par\end{center}

\noindent \begin{center}
$V^{U}=\left\{ v_{i}\in V|\neg\exists j:1\leqslant j\leqslant m,\mathbf{a}_{\mathbf{i}j}=1\right\} $
\par\end{center}

Finally, a definition for the neighborhood of each vertex is needed.
In this regard it must be noted that (1) this algorithm assumes directed
graphs, and (2) it only considers as neighbors of a person those people
related to the first one via a relationship started by that person. 

For instance, in a phone network those numbers a user makes calls
to would be neighbors, but not the numbers from which he receives
calls; in the blogosphere the neighborhood would comprise the blogs
a given blog links to, but not those linking to that blog; in Twitter
the neighbors would be those users a given user is following, but
not his followers. 

Thereby, the neighborhood for vertex $v_{i}\in V$ would be:

\noindent \begin{center}
$N_{i}=\left\{ v_{j}\in V|\exists e_{ij}\in E\right\} $
\par\end{center}

All of this defines the input graph but not the way in which the algorithm
works on it. As it has been said, it is an iterative algorithm and,
hence, at its core there is an operation to compute new weights for
each vertex attribute vector from its neighbors weights in the previous
iteration. It must be noted that only the weights for vertices belonging
to $V^{U}$ are updated, those from the originally labeled set $V^{K}$
are not assigned new weights:

\noindent \begin{center}
$\forall v_{i}\in V^{U}:\mathbf{a}_{\mathbf{i}}^{(t)}=\frac{1}{Z}\underset{v_{j}\in N_{i}}{\sum}\mathbf{a}{}_{\mathbf{j}}^{(t-1)}$
\par\end{center}

\noindent \begin{center}
$\forall v_{i}\in V^{K}:\mathbf{a_{i}}^{(t)}=\mathbf{a_{i}}^{(0)}$
\par\end{center}

In the previous formalization $Z$ is a normalizer. 

\emph{McC-Splat}, like other graph iterative algorithms, converges
after relatively few iterations. Hence, once weight vectors have stabilized
--or after a predefined number of iterations-- a large part of vertices
in $V^{U}$ have got weight vectors for the attribute of interest.
Other algorithms would then assign to each vertex the label with the
highest weight within the vector, or would require an \emph{ad hoc}
threshold to be defined for each class value. \emph{McC-Splat}, instead,
introduces two extra steps which can be used to achieve automatic
thresholding in a number of ways.

First of all, a fictitious sink vertex can be introduced. Such a vertex
would represent an individual related to every single person within
the social network. The weights for that vertex are computed after
the last iteration and they provide a measure of which weights could
be expected for a user without homophilous relationships. The usefulness
of such an approach is clear when a large majority of people belongs
to a single class; if that prevalence is not taken into account most
of the unknown individuals would be incorrectly assigned to the majority
class. This equation defines the weight vector for such a sink vertex:

\noindent \begin{center}
$\mathbf{s}^{(T+1)}=\frac{1}{Z}\underset{v_{i}\in V}{\sum}\mathbf{a}{}_{\mathbf{i}}^{(T)}$
\par\end{center}

Once the sink vertex weights are computed they can be used in two
ways: (1) vertices from $V^{U}$ can be assigned the label with the
highest weight which is \emph{also} above the corresponding weight
in the sink vector; or (2) vertices from $V^{U}$ can be assigned
the label with the weight which most largely departs --in percentage
value-- from the corresponding weight in the sink vector. 

The second approach to automatic thresholding requires to compute
an alternative weight vector for the members of $V^{K}$. As it has
been said, those vertices' vectors have got one single component with
a unity value --i.e. the component corresponding to the class each
individual belongs to-- and their vectors are not modified as the
algorithm iterates. However, it is possible to compute from their
corresponding neighborhoods the weights they would have whether they
had belonged to $V^{U}$:

\noindent \begin{center}
$\forall v_{i}\in V^{K}:\mathbf{a'}_{\mathbf{i}}^{(T+1)}=\frac{1}{Z}\underset{v_{j}\in N_{i}}{\sum}\mathbf{a}{}_{\mathbf{j}}^{(T)}$
\par\end{center}

By doing that it is possible to produce a reverse-ordered ranking
of individuals for each of the class values the attribute can take.
That way, instead of defining an \emph{ad hoc} threshold to decide
if a weight is high enough to accept the induced label, it is possible
to find different weights at different percentile values. 

Thereby, when using \emph{McC-Splat} it is not needed to take \emph{ad
hoc} weight thresholds; instead, the confidence required from the
labeled output can be chosen. For instance, by choosing the 90th percentile
only those members of $V^{U}$ whose weights were above 90\% of the
weights of $V^{K}$ members would appear in the output labeled set.

So, in short, \emph{McC-Splat} comes in the following flavors: (1)
\emph{Plain-vanilla}, the class with the highest weight is assigned%
\footnote{The \emph{unknown} class is ignored, otherwise all of the vertices
would remain unknown unless the number of labeled examples surpassed
the number of unknown vertices.%
}. (2) \emph{Sink-absolute}, the class with the highest weight and
above the corresponding weight within the sink node is assigned. (3)
\emph{Sink-relative}, the class with the highest positive difference
against the corresponding weight within the sink node is assigned.
(4) \emph{Percentile}, the class with the highest percentile --according
to the labeled individuals-- is assigned. Optionally, a minimum value
--e.g. 90\%-- can be forced or, otherwise, the \emph{unknown} class
is assigned.

Now, the algorithm's name should be self-explained: it is a multiclass
classifier which iteratively propagates weight vectors to every node
from its neighborhood; because each node's vector roughly represent
its likelihood of belonging to each class, the labeling is performed
in a {}``soft'' rather than in a {}``hard'' way; moreover, the
algorithm provides alternatives to automatically determine the most
reliable class for each node, making \emph{ad hoc }thresholds unnecessary.

\section{Experimental Evaluation}

\subsection{Dataset Description}

Social network data was needed to test the performance of \emph{McC-Splat}.
The graph depicting relationships between individuals was essential
but, in addition to that, a part of those users had to be labeled. 

Hence, the Twitter%
\footnote{Twitter is a microblogging and social networking service. Users publish
short text messages (\emph{tweets}) which are shown to all of their
followers. Relationships in Twitter are asymmetrical and, thus, a
user has got followers and \emph{followees}.%
} dataset collected in \cite{NepotisticGayo} was used. It comprises
27.9 million English-written tweets published from January 26 to August
31, 2009 by 4.98 million users. 

Followers and followees for each of the users in that dataset were
also collected. Links to users not appearing in the dataset were disregarded,
and isolated users were removed. Furthermore, a substantial amount
of user accounts were suspended at the moment of the graph crawl and,
hence, no information on them was available. Lastly, because of the
unavoidable network problems, coupled with the fact that the API was
pushed a little too far, the information for a noticeable amount of
users was not eventually crawled. 

Thus, the user graph consisted of 1.8 million users with their corresponding
links and profiles --i.e. full name, short biography, location, etc.
Given that at the moment of collecting the dataset, the number of
Twitter users in the U.S. was estimated between 14 and 18 millions%
\footnote{\texttt{http://www.socialtimes.com/2009/04/twitter-14-million/}, \texttt{http://mashable.com/2009/09/14/twitter-2009-stats/}%
}, and that most of the crawled users were supposed to be from the
U.S. it can be considered a rather substantial sample.

\subsection{Labeling Twitter users}

Unlike other OSNs such as Facebook, Twitter profiles do not provide
highly structured information; there is no way, for instance, to indicate
the user's sex or age. Instead, Twitter profiles consist of the user's
full name, location, website, and a short biography. All of these
fields are free text and there is a high disparity in their use. For
example, 62.31\% of the users in the dataset provide a location string,
but only 36.46\% provide their full personal name \cite{NepotisticGayo}.

This does not mean that no personal information can be extracted from
Twitter profiles. Quite to the contrary, using the location, full
name, and biography strings, half of the users in the dataset were
geolocated, the sex of one third of them was found, in addition to
the age for about 11,000 \cite{NepotisticGayo}. Needless to say,
the data was noisy, and the labeling methods a bit rough; though,
anecdotal evidence revealed a quite accurate big-picture of Twitter
demographics. 

Therefore, a similar approach was employed to label users according
to a number of personal traits. In addition to sex and age, the following
attributes%
\footnote{All of the chosen attributes, except for \emph{race/ethnicity}, appear
in Facebook profiles and, thus, it would not be surprising to find
information on them in Twitter biographies.%
} were also chosen: \emph{political orientation}, \emph{religious affiliation},
\emph{race and ethnicity}, and \emph{sexual orientation}. All of them
are usually considered sensitive information, and most countries have
enacted laws against discrimination based on any of such attributes.
In spite of this, many people still feel the need to hide those personal
details. Thus, it is important to find out the degree in which such
individuals can be inadvertently exposed because of their acquaintances.

{\scriptsize }%
\begin{table*}
{\scriptsize \caption{Classes for each of the six personal attributes along the rules applied
to label Twitter users according to them. All of the labels, except
for \emph{sex} were obtained by pattern-matching the users' biographies.
The age intervals were those used by \cite{WWW07Demographics}.}
\label{tabla1}}{\scriptsize \par}

\noindent \centering{}{\scriptsize }\begin{tabular}{|>{\centering}m{2.5cm}|c|>{\centering}m{1\columnwidth}|c|}
\hline 
\textbf{\scriptsize Attribute} & \textbf{\scriptsize Class} & \textbf{\scriptsize Rule or pattern} & \textbf{\scriptsize \# users}\tabularnewline
\hline
\hline 
{\scriptsize \multirow{2}{2.5cm}{sex}} & {\scriptsize female} & \centering{}{\scriptsize \multirow{2}{1\columnwidth}{User name had to be composed of first and last name from the U.S. Census. Sex was assigned according to frequency of use of the first name in U.S. population.}} & {\scriptsize 271,539}\tabularnewline
\cline{2-2} \cline{4-4} 
 & {\scriptsize male} &  & {\scriptsize 384,574}\tabularnewline
\hline 
{\scriptsize \multirow{5}{2.5cm}{age}} & {\scriptsize teenage} & {\scriptsize \multirow{5}{1\columnwidth}{Age was extracted from the user's bio looking for the patterns \emph{year-old} or \emph{years old} preceded by a number or a numeral. Then, ages <18, 18-24, 25-34, 35-49, and >49 were assigned to each class.}} & {\scriptsize 3,483}\tabularnewline
\cline{2-2} \cline{4-4} 
 & {\scriptsize youngster} &  & {\scriptsize 4,562}\tabularnewline
\cline{2-2} \cline{4-4} 
 & {\scriptsize young} &  & {\scriptsize 1,911}\tabularnewline
\cline{2-2} \cline{4-4} 
 & {\scriptsize mid-age} &  & {\scriptsize 663}\tabularnewline
\cline{2-2} \cline{4-4} 
 & {\scriptsize elder} &  & {\scriptsize 296}\tabularnewline
\hline 
{\scriptsize \multirow{2}{2.5cm}{political orientation}} & {\scriptsize democrat} & \textbf{\scriptsize democrat{*}}{\scriptsize , lib-dem{*}, libdem{*},
dems{*}} & {\scriptsize 248}\tabularnewline
\cline{2-4} 
 & {\scriptsize republican} & {\scriptsize conservat{*}, gop, g.o.p., palin, pro-life{*}, prolife{*},
}\textbf{\scriptsize republican{*}}{\scriptsize , right-wing, rightish,
tcot, tea-party, teaparty} & {\scriptsize 2,040}\tabularnewline
\hline 
{\scriptsize \multirow{5}{2.5cm}{religious affiliation}} & {\scriptsize atheist} & {\scriptsize agnost{*}, anti-theis{*}, antitheis{*}, ateus, }\textbf{\scriptsize atheis{*}}{\scriptsize ,
athiest{*}, empiricist{*}, godless{*}, heathen{*}, humanism{*}, humanist{*},
irreligion{*}, non-believer{*}, non-theist{*}, nonbeliever{*}, nontheist{*},
pagan{*}, rational{*}, sceptic{*}, secular{*}, skepchicks, skeptic{*}} & {\scriptsize 330}\tabularnewline
\cline{2-4} 
 & {\scriptsize buddhist} & \textbf{\scriptsize buddh{*}}{\scriptsize , dhamma{*}, dharma{*},
sangha, twangha, vipassana, yoga{*}, yoginis, yogis, zen} & {\scriptsize 204}\tabularnewline
\cline{2-4} 
 & {\scriptsize christian} & {\scriptsize adventist{*}, anglican{*}, baptist{*}, cathol{*}, cattolici,
}\textbf{\scriptsize christ{*}}{\scriptsize , church{*}, evangelical,
gospel{*}, jesus{*}, lutheran{*}, methodist{*}, minister{*}, ministries{*},
ministry{*}, pastor{*}, pentecostal{*}, preacher{*}, presbyterian{*},
priest{*}} & {\scriptsize 8,103}\tabularnewline
\cline{2-4} 
 & {\scriptsize jewish} & {\scriptsize circumcision, israel{*}, jerusalem, jew, }\textbf{\scriptsize jewish}{\scriptsize ,
jews, judaism, jude, kosher, rabbi, sephardic, synagogue{*}, torah,
yiddish, zion{*}} & {\scriptsize 458}\tabularnewline
\cline{2-4} 
 & {\scriptsize muslim} & {\scriptsize imam, islam{*}, isulamic{*}, mosque{*}, }\textbf{\scriptsize muslim{*}}{\scriptsize ,
quran, salaam, tweeplims} & {\scriptsize 171}\tabularnewline
\hline 
{\scriptsize \multirow{6}{2.5cm}{race/ethnicity}} & {\scriptsize asian-american} & \textbf{\scriptsize asian{*}}{\scriptsize , chinese-american, filipin{*},
hindu{*}, india, indian-american, japan, japanese-american, korea{*},
taoism, vietnam{*}} & {\scriptsize 65}\tabularnewline
\cline{2-4} 
 & {\scriptsize black} & {\scriptsize africa{*}, }\textbf{\scriptsize black}{\scriptsize ,
black-american, black-man, black-woman, hip-hop{*}, hiphop{*}} & {\scriptsize 202}\tabularnewline
\cline{2-4} 
 & {\scriptsize hispanic} & {\scriptsize amigo, belleza, familia, favoritos, gente, }\textbf{\scriptsize hispanic}{\scriptsize ,
latina, latino, mexico} & {\scriptsize 6}\tabularnewline
\cline{2-4} 
 & {\scriptsize native-american} & {\scriptsize aboriginal, alaska-native, american-indian, first-nation,
firstnation, indigenous{*}, native american, }\textbf{\scriptsize native-american} & {\scriptsize 80}\tabularnewline
\cline{2-4} 
 & {\scriptsize native-hawaiian} & {\scriptsize aloha, hawaii{*}, honolulu, native hawaiian, }\textbf{\scriptsize native-hawaiian}{\scriptsize ,
oahu, ohana} & {\scriptsize 4}\tabularnewline
\cline{2-4} 
 & {\scriptsize white} & {\scriptsize caucasian, }\textbf{\scriptsize white}{\scriptsize ,
white-american, white-man, white-woman} & {\scriptsize 24}\tabularnewline
\hline 
{\scriptsize \multirow{2}{2.5cm}{sexual orientation}} & {\scriptsize heterosexual} & \textbf{\scriptsize hetero{*}} & {\scriptsize 15}\tabularnewline
\cline{2-4} 
 & {\scriptsize homosexual} & {\scriptsize bisexual{*}, gay{*}, glbt, glsen, gltb, homo-{*}, }\textbf{\scriptsize homosex{*}}{\scriptsize ,
l-word, lesbian{*}, lgbt, lgbtq, marriage-equality, queer, transgender} & {\scriptsize 1,471}\tabularnewline
\hline
\end{tabular}
\end{table*}
{\scriptsize \par}

All of the classes, except those corresponding to sex, were determined
by means of pattern matching (see table \ref{tabla1} for the patterns
applied). Firstly, each class name was used to obtain a initial list
of users. For instance, the patterns \texttt{democrat{*}} and \texttt{republican{*}}
were used to find users self-defined as Democrats or Republicans.
Once there was a preliminary list of users for each class, their biographies
were mined to find the most frequent keywords which could be considered
indicative of class belonging. That way, for example, patterns such
as \texttt{lib-dem{*}} or \texttt{dems{*}} were found for Democrats,
and \texttt{conservat{*}} or \texttt{tea party} for Republicans.

Certainly, such a labeling method is error prone but the goal was
to obtain the largest%
\footnote{Table \ref{tabla1} reveals that the number of labeled examples obtained
was rather low for all of the attributes and, unsurprisingly, the
more sensitive the attribute, the fewer users disclose information
about it in their biographies.%
} possible labeled set for each class and attribute. Because of the
nature of \emph{McC-Splat}, it was assumed that large although noisy
data was preferable to cleaner but small samples. After all, should
the results be encouraging, better labeling approaches could be used.

Finally, the labeled sets were split into \emph{training} and \emph{tests}
partitions: the former consisted of a random selection of 80\% of
the users in each class and was used as input for the algorithm; the
later comprised the remaining 20\% of the users and was left out for
evaluation.

\subsection{Results}

\emph{McC-Splat} was applied to the Twitter graph in each of its four
different {}``flavors'' just considering the users in the training\emph{
}partitions. That way, labels were obtained for the rest of the users
in the graph including those in the test partitions. Then, by comparing
the algorithm's class assignments for those users with the actual
class belonging according to their biographies, precision and recall
figures were computed (see table \ref{TMcC-SplatPerformance}). For
comparison purposes, the performance of a random classifier based
on the proportion of each of the different classes is shown in table
\ref{TRandomClassifier}.

In addition to that first experiment, a second one was conducted on
another independently labeled set. To that end, data was collected
from \emph{WeFollow}%
\footnote{\texttt{http://wefollow.com}%
} which is a Twitter user directory where users classify themselves
according to the topics they are interested in. Each topic is represented
by a tag, and a list of users following each tag can be obtained%
\footnote{For instance, \texttt{http://wefollow.com/twitter/democrat} gives
access to a list of users self-defined as Democrat, while \texttt{http://wefollow.com/twitter/republican}
provides a list of Republican users.%
}. 

Hence, most of the patterns from table \ref{tabla1} were employed
to obtain lists of users from WeFollow%
\footnote{Sex and age were not able to be tested with data from WeFollow.%
}. Needless to say, not every user in those lists appeared in the Twitter
user graph and, therefore, those users not appearing in the graph,
in addition to those already labeled --i.e. appearing in the training
and test partitions-- were removed. 

Performance results on this second dataset for both the random classifier
and the \emph{McC-Splat} algorithm can be seen in tables \ref{RandomClassifierWeFollow}
and \ref{TMcCSplatWeFollow}, respectively.

{\scriptsize }%
\begin{table*}
{\scriptsize \caption{Performance of a random classifier based on the proportion of each
class in the labeled data and working on the same labeled data.}
\label{TRandomClassifier}}{\scriptsize \par}

\centering{}{\scriptsize }\begin{tabular}{|c|c|c|c|c|}
\hline 
\textbf{\scriptsize Attribute} & \multicolumn{2}{c|}{\textbf{\scriptsize P=R=}{\scriptsize $F_{1}$}} & \textbf{\scriptsize Class} & \multicolumn{1}{c|}{\textbf{\scriptsize P=R=}{\scriptsize $F_{1}$}}\tabularnewline
\hline 
{\scriptsize \multirow{2}{*}{sex}} & {\scriptsize Micro-avg.} & {\scriptsize 0.5148} & {\scriptsize female} & {\scriptsize 0.4139}\tabularnewline
 & {\scriptsize Macro-avg.} & {\scriptsize 0.5} & {\scriptsize male} & {\scriptsize 0.5861}\tabularnewline
\hline 
{\scriptsize \multirow{5}{*}{age}} & {\scriptsize \multirow{2}{*}{Micro-avg.}} & {\scriptsize \multirow{2}{*}{0.3116}} & {\scriptsize teenage} & {\scriptsize 0.3191}\tabularnewline
 &  &  & {\scriptsize youngster} & {\scriptsize 0.4180}\tabularnewline
 & {\scriptsize \multirow{3}{*}{Macro-avg.}} & {\scriptsize \multirow{3}{*}{0.2}} & {\scriptsize young} & {\scriptsize 0.1751}\tabularnewline
 &  &  & {\scriptsize mid-age} & {\scriptsize 0.0607}\tabularnewline
 &  &  & {\scriptsize elder} & {\scriptsize 0.0271}\tabularnewline
\hline 
{\scriptsize \multirow{5}{*}{religious affiliation}} & {\scriptsize \multirow{2}{*}{Micro-avg.}} & {\scriptsize \multirow{2}{*}{0.7693}} & {\scriptsize atheist} & {\scriptsize 0.0356}\tabularnewline
 &  &  & {\scriptsize budhist} & {\scriptsize 0.0220}\tabularnewline
 & {\scriptsize \multirow{3}{*}{Macro-avg.}} & {\scriptsize \multirow{3}{*}{0.2}} & {\scriptsize christian} & {\scriptsize 0.8745}\tabularnewline
 &  &  & {\scriptsize jewish} & {\scriptsize 0.0494}\tabularnewline
 &  &  & {\scriptsize muslim} & {\scriptsize 0.0185}\tabularnewline
\hline 
{\scriptsize \multirow{2}{*}{political orientation}} & {\scriptsize Micro-avg.} & {\scriptsize 0.8068} & {\scriptsize democrat} & {\scriptsize 0.1084}\tabularnewline
 & {\scriptsize Macro-avg.} & {\scriptsize 0.5} & {\scriptsize republican} & {\scriptsize 0.8916}\tabularnewline
\hline 
{\scriptsize \multirow{2}{*}{sexual orientation}} & {\scriptsize Micro-avg.} & {\scriptsize 0.9798} & {\scriptsize heterosexual} & {\scriptsize 0.0101}\tabularnewline
 & {\scriptsize Macro-avg.} & {\scriptsize 0.5} & {\scriptsize homosexual} & {\scriptsize 0.9899}\tabularnewline
\hline 
{\scriptsize \multirow{6}{*}{race/ethnicity}} & {\scriptsize \multirow{3}{*}{Micro-avg.}} & {\scriptsize \multirow{3}{*}{0.3586}} & {\scriptsize asian-american} & {\scriptsize 0.1706}\tabularnewline
 &  &  & {\scriptsize black} & {\scriptsize 0.5302}\tabularnewline
 &  &  & {\scriptsize hispanic} & {\scriptsize 0.0157}\tabularnewline
 & {\scriptsize \multirow{3}{*}{Macro-avg.}} & {\scriptsize \multirow{3}{*}{0.1667}} & {\scriptsize native-american} & {\scriptsize 0.2100}\tabularnewline
 &  &  & {\scriptsize native-hawaiian} & {\scriptsize 0.0105}\tabularnewline
 &  &  & {\scriptsize white} & {\scriptsize 0.0630}\tabularnewline
\hline
\end{tabular}
\end{table*}
{\scriptsize }%
\begin{table*}
{\scriptsize \caption{Performance of a random classifier based on the proportion of each
class in the labeled data and working on the WeFollow dataset.}
\label{RandomClassifierWeFollow}}{\scriptsize \par}

\centering{}{\scriptsize }\begin{tabular}{|c|c|ccc|c|ccc|}
\hline 
\textbf{\scriptsize Attribute} &  & \textbf{\scriptsize P} & \textbf{\scriptsize R} & {\scriptsize $F_{1}$} & \textbf{\scriptsize Class} & \multicolumn{1}{c}{\textbf{\scriptsize P}} & \textbf{\scriptsize R} & {\scriptsize $F_{1}$}\tabularnewline
\hline 
{\scriptsize \multirow{5}{*}{religious affiliation}} & {\scriptsize \multirow{2}{*}{Micro-avg.}} & \multicolumn{3}{c|}{{\scriptsize \multirow{2}{*}{0.6843}}} & {\scriptsize atheist} & {\scriptsize 0.0872} & {\scriptsize 0.0356} & {\scriptsize 0.0506}\tabularnewline
 &  &  &  &  & {\scriptsize budhist} & {\scriptsize 0.0513} & {\scriptsize 0.0220} & {\scriptsize 0.0308}\tabularnewline
 & {\scriptsize \multirow{3}{*}{Macro-avg.}} & \multicolumn{3}{c|}{{\scriptsize \multirow{3}{*}{0.2}}} & {\scriptsize christian} & {\scriptsize 0.7741} & {\scriptsize 0.8745} & {\scriptsize 0.8213}\tabularnewline
 &  &  &  &  & {\scriptsize jewish} & {\scriptsize 0.0484} & {\scriptsize 0.0494} & {\scriptsize 0.0489}\tabularnewline
 &  &  &  &  & {\scriptsize muslim} & {\scriptsize 0.0389} & {\scriptsize 0.0185} & {\scriptsize 0.0250}\tabularnewline
\hline 
{\scriptsize \multirow{2}{*}{political orientation}} & {\scriptsize Micro-avg.} & \multicolumn{3}{c|}{{\scriptsize 0.7966}} & {\scriptsize democrat} & {\scriptsize 0.1213} & {\scriptsize 0.1084} & {\scriptsize 0.1145}\tabularnewline
 & {\scriptsize Macro-avg.} & \multicolumn{3}{c|}{{\scriptsize 0.5}} & {\scriptsize republican} & {\scriptsize 0.8787} & {\scriptsize 0.8916} & {\scriptsize 0.8851}\tabularnewline
\hline 
{\scriptsize \multirow{2}{*}{sexual orientation}} & {\scriptsize Micro-avg.} & \multicolumn{3}{c|}{{\scriptsize 0.9899}} & {\scriptsize heterosexual} & {\scriptsize 0} & {\scriptsize 1} & {\scriptsize 0}\tabularnewline
 & {\scriptsize Macro-avg.} & {\scriptsize 0.5} & {\scriptsize 0.9950} & {\scriptsize 0.6655} & {\scriptsize homosexual} & {\scriptsize 1} & {\scriptsize 0.9899} & {\scriptsize 0.9949}\tabularnewline
\hline 
{\scriptsize \multirow{6}{*}{race/ethnicity}} & {\scriptsize \multirow{3}{*}{Micro-avg.}} & \multicolumn{3}{c|}{{\scriptsize \multirow{3}{*}{0.2119}}} & {\scriptsize asian-american} & {\scriptsize 0.6022} & {\scriptsize 0.1706} & {\scriptsize 0.2659}\tabularnewline
 &  &  &  &  & {\scriptsize black} & {\scriptsize 0.1853} & {\scriptsize 0.5302} & {\scriptsize 0.2746}\tabularnewline
 &  &  &  &  & {\scriptsize hispanic} & {\scriptsize 0.1735} & {\scriptsize 0.0157} & {\scriptsize 0.0289}\tabularnewline
 & {\scriptsize \multirow{3}{*}{Macro-avg.}} & {\scriptsize \multirow{3}{*}{0.1667}} & {\scriptsize \multirow{3}{*}{0.4878}} & {\scriptsize \multirow{3}{*}{0.2484}} & {\scriptsize native-american} & {\scriptsize 0.0391} & {\scriptsize 0.2100} & {\scriptsize 0.0659}\tabularnewline
 &  &  &  &  & {\scriptsize native-hawaiian} & {\scriptsize 0} & {\scriptsize 1} & {\scriptsize 0}\tabularnewline
 &  &  &  &  & {\scriptsize white} & {\scriptsize 0} & {\scriptsize 1} & {\scriptsize 0}\tabularnewline
\hline
\end{tabular}
\end{table*}
{\scriptsize \par}

\subsection{Discussion of Results}

As it can be seen from tables \ref{TMcC-SplatPerformance} and \ref{TMcCSplatWeFollow}
the performance of \emph{McC-Splat} was notably high. Average precision
and accuracy figures were quite similar, implying that performance
across classes within the same attribute is comparable and, thus,
there was no much bias towards the prevalent classes. 

Attributes such as \emph{religious affiliation}, \emph{political orientation},
\emph{sexual orientation}, and \emph{race and ethnicity} achieved
above 95\% precision when evaluating on the test partitions. Results
in the WeFollow dataset were very similar, except for \emph{race/ethnicity}
where precision dropped to 50\% and accuracy to 71\%. 

The poorest results were achieved when assigning \emph{sex} and \emph{age}:
62\% and 43\% macro-averaged precision, respectively. With regards
to \emph{age}, maybe it was problematic because it is actually a continuous
variable. After reviewing the actual classifications it was found
that most of the errors were due to assigning users to nearby classes
--e.g. classifying \emph{teenagers} as \emph{youngsters}, \emph{youngsters}
as \emph{youngs}, etc.

All in all, \emph{McC-Splat} clearly outperformed the random classifier
by an exceedingly large margin although, certainly, when an attribute
has got a clearly prevalent class it is much more difficult to outperform
it. In the presence of such prevalent classes the random classifier
achieved good accuracy but also poor macro-averaged precision; \emph{McC-Splat},
instead, was not very affected by such prevalent classes and it exhibited
comparable precision across classes.

Regarding the different {}``flavors'', the \emph{Plain-vanilla}
version did not outperform the random classifier for prevalent classes
(e.g. \emph{male} vs \emph{female}, \emph{young} vs the rest of \emph{age}
intervals, and \emph{christians} vs the rest of \emph{religious affiliations}),
and it even underperformed when classifying \emph{homosexual} individuals.
The rest of the {}``flavors'' clearly outperformed the random classifier
--even for prevalent classes-- and they consistently achieved high
performance figures. Therefore, \emph{Plain-vanilla} could be disregarded
and additional experiments are required to find which of the other
three alternatives can be the best choice. In this regard, better
labeled data --in particular for large majority classes-- is also
needed.

\section{Related Work}

As it has been said, \emph{McC-Splat} is a semi-supervised graph labeling
algorithm based on label propagation. There are other algorithms which
are somewhat similar and, hence, those most highly related are to
be briefly reviewed. 

Maybe the best known iterative graph algorithm is PageRank \cite{PageRank1998},
it computes for each vertex --generally a web page-- a score which
corresponds to its relevance within the network. Its popularity has
spurred the use of similar methods in many other scenarios --e.g.
to fight spam in the Web \cite{TrustRank2004}.

With regards to the use of the graph structure to perform classification,
one of the earliest works was a hypertext classifier \cite{Chakrabarti1998}.
In this case, however, the links were used to improve the classifier
but other clues --such as the documents content-- were also required. 

Much more related to \emph{McC-Splat} are the works described in \cite{Macskassy2003,Neville2000}.
In \cite{Neville2000} it is described an iterative application of
Bayesian classifiers where the objects attributes were modified from
the inferences made on their neighbors in each iteration. In \cite{Macskassy2003}
the so-called wvRN%
\footnote{Weighted-vote Relational Network classifier.%
} method is described. That algorithm works on undirected weighted
graphs and just relies on the objects labels and relationships. It
estimates the probability of an object belonging to a given class
as the weighted proportion of its neighbors that belong to that class
and, then, the majority label is assigned after each iteration. 

Although related, there are several differences between wvRN and \emph{McC-Splat}:
the later works on unweighted directed graphs, labels are not assigned
by majority vote but, instead, weight vectors are propagated. Besides,
in the absence of labeled neighbors wvRN assigns label on the basis
of the class priors --i.e. a random classifier-- while \emph{McC-Splat}
assigns the \emph{unknown} class. Finally, the use of a sink node
and the estimation of weight vectors for the labeled examples to perform
auto-thresholding are novel additions which could be compared to cautious
classification \cite{McDowell2009}.

As it has been said, data mining users' relationships in OSNs raises
some concerns and, in fact, this study have exposed the privacy risks
due to the public nature of those relationships. Hence, this work
has got some points of similarity with a number of recent studies
on privacy in OSNs. 

It has been shown, for instance, that different kind of attacks can
be conducted on the basis of known relationships and group memberships
\cite{Zheleva2008,Zheleva2009}, and a number of studies have provided
additional support for those findings in Facebook --e.g. \cite{Gaydar2009,Lindamood2009}.

It has been stated that privacy attacks can be successful when \emph{{}``as
much as half of the profiles are private''} \cite{Zheleva2008}.
However, this study has revealed that the number of required known
users is, in fact, much lower --well below 1\% for a sample of 1.8
million users-- and the achieved precision is much higher than the
one reported in \cite{Zheleva2008}. Thereby, privacy issues because
of publicizing acquaintances in OSNs should be a major concern for
their users.

Finally, a few pertinent works on measures to improve privacy in OSNs
are referenced to provide context for the protocol described in the
last section.

At least two different Facebook applications relying on public key
cryptography to store obfuscated information in the OSN servers have
been proposed \cite{flybynight2008,FaceCloak2009}. By doing that
users can still make use of the OSN services but their personal information
is decrypted on the client side and, thus, it is inaccessible for
the OSN operator. Needless to say, encrypted text is relative easy
to detect and, thus, a {}``hostile'' OSN operator could disable
accounts using such a measure. 

Because of that, it has been proposed to use instead a dictionary
known to the members of a group \cite{NOYB2008}. Such a dictionary
would provide a way to replace {}``atoms'' of personal information
with atoms from other users. For instance, the name, age, or sex of
a user would be stored in such a way that they still resemble personal
information but cannot be linked to the actual individual. By using
the dictionary, the group members could translate that fake information
into the actual attributes of their acquaintance. Purportedly, this
measure is much more difficult to detect than cryptography and, thereby,
it could be applied even when using the services provided by {}``hostile''
OSN operators \cite{NOYB2008}.

\section{Implications and Conclusions}

\subsection{Implications for Users Privacy}

Users sensitive information, such as political or religious beliefs,
race and ethnicity, or sexual orientation can be determined with notable
precision from their neighbors with rather simple algorithms. Thereby,
it does not matter if users do not self-disclose personal traits,
they can be inadvertently exposed because of acquaintances who do
not conceal such information. 

Most works on privacy in OSNs have mainly focused on ways to guarantee
that released datasets do not put at risk the users' privacy --e.g.
\cite{Ying2009}. Certainly, such anonymization measures may dispel
some concerns the operators of OSNs can have about releasing data
for research purposes. However, it is not at all necessary to obtain
the data from the operator of the OSN, but it is relatively easy to
collect using the available APIs. Therefore, in spite of anonymization
methods, users of OSNs are fully exposed to any third party aiming
to data mine social graphs.

A trivial solution for that problem would be, of course, to disable
the APIs. This, however, is unlikely to happen because it would be
contrary to the interests of the operators of the OSNs. In addition
to that, it would just make difficult%
\footnote{Several ways in which an attacker can obtain information on network
relationships by compromising a number of user accounts are described
in \cite{Korolova2008}.%
} for third parties to mine the users data but would not prevent the
operator of the OSN and licensed third parties from doing it.

A number of works, some of them referenced in the previous section,
propose users to encrypt the information they submit to the system.
Needless to say, making the users information opaque for the OSN would
put at risk their current business models which, to a great or lesser
extent, revolve around marketing and personalization. Thereby, it
does not seem unreasonable to assume that if encryption went mainstream
among OSN users, the operators of the services would force users to
use plain text.

\subsection{Minimizing Data Mining Risks}

So, to sum up, graph anonymization is an unreliable passive%
\footnote{Passive, that is, from the point of view of the users.%
} measure, and heavy use of cryptography, an active user's measure,
could be easily disallowed by the operators of the OSNs. Hence, procedures
to minimize privacy leakages should be active and keep the use of
cryptography to a minimum; some hints on such a prophylactic protocol
are provided here.

First of all, the following protocol has been devised for asymmetrical
social networks in general, and Twitter in particular. Secondly, users
are responsible for the information they disclose on themselves; that
is, the purpose of this protocol is not to protect their privacy regardless
of their actions, but to minimize the likelihood of being exposed
because of their relationships. In third place, users cannot control
who is following them but who they follow. It has been shown that
these relationships are risky and, thus, identifiable accounts cannot
be used to follow anybody.

Needless to say, the network is useless if users are isolated and,
thereby, they need a mechanism to follow other users. To that end,
a second account is to be used. The nickname should be a totally random
string, and no information should be provided other than a public
key. This \emph{anonymous} account --in contrast to the previous \emph{identified}
account-- would not be used to post messages other than mentions to
followees, and it would not accept followers.

Obviously, using two different accounts would be pointless if they
can be linked to each other by means of the IP address. Therefore,
the anonymous account should connect to the service through an anonymizing
service such as \emph{Tor}%
\footnote{\texttt{https://www.torproject.org/}%
} or \emph{I2P}%
\footnote{\texttt{http://www.i2p2.de/}%
} while this is not necessary for the identified account.

With regards to message publishing, those not mentioning any account
or mentioning an identified account could be published unencrypted.
After all, users are responsible for what they publish on themselves,
and cannot control the messages other users address to them. However,
if the message is a reply from an identified account to an anonymous
account it should be fully encrypted using the public key corresponding
to the anonymous account. The reason for this is to avoid eavesdroppers
to find out implicit links starting on identified accounts.

The most cumbersome part would be the one regarding the exchange of
\emph{credentials} between anonymous and identified accounts. Such
an exchange would be needed to allow users to follow their followers.
As it has been said, anonymous accounts are not for publishing messages
and, thus, they would be of no interest. However, after receiving
a new follower, that anonymous account is the only piece of information
the user has got to reach the follower's identified account. Hence,
the user receiving a new follower should publish his or her public
key encrypted with the public key of the new follower. The follower
would publish, in return, the nickname for his or her identified account
encrypted with the public key of the followee. At that point, the
followee could use his anonymous account to start following the identified
account of his new follower. 

Clearly, that chain of actions would allow an eavesdropper to link
anonymous and identified accounts. Thus, to avoid it, the exchange
of credentials could be made at pre-scheduled hours. In addition to
this, it must be clear that this protocol does not aim to maintain
users anonymous from each other but to conceal their relationships
from third parties observing the social network --including its operators.

Finally, all of these measures should be implemented by client software
in such a way that the user could use the OSN transparently.

\subsection{Final Remarks and Future Work}

A new algorithm to perform user profiling in social networks, \emph{McC-Splat},
has been described. The new method is related to other known algorithms
but, unlike them, it does not require \emph{ad hoc} thresholds but,
instead, it provides a number of alternatives to perform auto thresholding
from the input labeled data.

A number of experiments were conducted to test its performance. Results
from those experiments have been reported, revealing that \emph{McC-Splat}
largely outperforms a random classifier and, in fact, achieves a notably
high precision for very different classes and attributes. Nevertheless,
further experiments are needed to determine which of the different
{}``flavors'' of the algorithm is the best choice, in addition to
test the algorithm on data from OSNs other than Twitter, and labeled
by different means.

The attributes employed for the experiments are usually considered
sensitive personal information and, thus, the experiments had an additional
outcome: exposing the risk that acquaintances suppose for users which
can be exposed even without revealing any personal information on
themselves.

Thereby, a prophylactic protocol to minimize leakages due to graph
data mining was outlined. Further work is needed in this regard: a
prototype implementation is highly needed; in addition to field studies
regarding its use by real users, and analyzing its sensitiveness to
different kind of attacks --mainly those based on infiltration.

\bibliographystyle{plain}
\bibliography{icwsm-2011}

\noindent \begin{center}
{\scriptsize }%
\begin{sidewaystable*}
{\scriptsize \caption{Performance figures for the six attributes and the four different
{}``flavors'' of the \emph{McC-Splat} algorithm working on the Twitter
dataset. Details for each individual class are provided in addition
to aggregated figures: both micro- and macro-averaged. Micro-averaged
precision is equivalent to the accuracy of the classifier for each
attribute. Figures in bold correspond to {}``material'' performance
improvements against the random classifier --i.e. larger than 10\%,
according to the criterion proposed by \cite{SparckJones1974}. }
\label{TMcC-SplatPerformance}}{\scriptsize \par}

{\scriptsize }\begin{tabular}{|c|c|ccc|ccc|ccc|ccc|}
\hline 
{\scriptsize \multirow{2}{*}{}\textbf{\scriptsize Attribute}{\scriptsize }} & {\scriptsize \multirow{2}{*}{}\textbf{\scriptsize Class}{\scriptsize }} & \multicolumn{3}{c|}{\textbf{\scriptsize Plain-vanilla}} & \multicolumn{3}{c|}{\textbf{\scriptsize Sink-absolute}} & \multicolumn{3}{c|}{\textbf{\scriptsize Sink-relative}} & \multicolumn{3}{c|}{\textbf{\scriptsize Percentile}}\tabularnewline
\cline{3-14} 
 &  & {\scriptsize P} & {\scriptsize R} & {\scriptsize $F_{1}$} & {\scriptsize P} & {\scriptsize R} & {\scriptsize $F_{1}$} & {\scriptsize P} & {\scriptsize R} & {\scriptsize $F_{1}$} & {\scriptsize P} & {\scriptsize R} & {\scriptsize $F_{1}$}\tabularnewline
\hline
\hline 
{\scriptsize \multirow{4}{*}{sex}} & {\scriptsize female} & \textbf{\scriptsize 0.6307} & {\scriptsize 0.1091} & {\scriptsize 0.186} & \textbf{\scriptsize 0.6149} & {\scriptsize 0.3135} & {\scriptsize 0.4153} & \textbf{\scriptsize 0.6125} & {\scriptsize 0.3356} & {\scriptsize 0.4337} & \textbf{\scriptsize 0.6174} & {\scriptsize 0.2432} & {\scriptsize 0.3489}\tabularnewline
 & {\scriptsize male} & {\scriptsize 0.6039} & {\scriptsize 0.8852} & {\scriptsize 0.718} & \textbf{\scriptsize 0.6803} & {\scriptsize 0.4792} & {\scriptsize 0.5623} & \textbf{\scriptsize 0.6907} & {\scriptsize 0.4679} & {\scriptsize 0.5579} & \textbf{\scriptsize 0.6765} & {\scriptsize 0.3318} & {\scriptsize 0.4452}\tabularnewline
\cline{2-14} 
 & {\scriptsize Macro-avg.} & \textbf{\scriptsize 0.6173} & {\scriptsize 0.4972} & {\scriptsize 0.5507} & \textbf{\scriptsize 0.6476} & {\scriptsize 0.3964} & {\scriptsize 0.4917} & \textbf{\scriptsize 0.6516} & {\scriptsize 0.4018} & {\scriptsize 0.497} & \textbf{\scriptsize 0.6469} & {\scriptsize 0.2875} & {\scriptsize 0.3981}\tabularnewline
 & {\scriptsize Micro-avg} & \textbf{\scriptsize 0.606} & {\scriptsize 0.564} & {\scriptsize 0.5842} & \textbf{\scriptsize 0.6582} & {\scriptsize 0.4106} & {\scriptsize 0.5057} & \textbf{\scriptsize 0.6623} & {\scriptsize 0.4132} & {\scriptsize 0.5089} & \textbf{\scriptsize 0.6551} & {\scriptsize 0.2951} & {\scriptsize 0.4069}\tabularnewline
\hline 
{\scriptsize \multirow{7}{*}{age}} & {\scriptsize teenage} & \textbf{\scriptsize 0.533} & {\scriptsize 0.1392} & {\scriptsize 0.2207} & \textbf{\scriptsize 0.5112} & {\scriptsize 0.1636} & {\scriptsize 0.2478} & \textbf{\scriptsize 0.5398} & {\scriptsize 0.175} & {\scriptsize 0.2644} & \textbf{\scriptsize 0.5989} & {\scriptsize 0.1564} & {\scriptsize 0.248}\tabularnewline
 & {\scriptsize youngster} & {\scriptsize 0.4438} & {\scriptsize 0.8697} & {\scriptsize 0.5877} & \textbf{\scriptsize 0.5} & {\scriptsize 0.2267} & {\scriptsize 0.312} & \textbf{\scriptsize 0.5375} & {\scriptsize 0.1961} & {\scriptsize 0.2873} & \textbf{\scriptsize 0.5464} & {\scriptsize 0.1742} & {\scriptsize 0.2641}\tabularnewline
 & {\scriptsize young} & \textbf{\scriptsize 0.3607} & {\scriptsize 0.0574} & {\scriptsize 0.0991} & \textbf{\scriptsize 0.2825} & {\scriptsize 0.1305} & {\scriptsize 0.1786} & \textbf{\scriptsize 0.2458} & {\scriptsize 0.1149} & {\scriptsize 0.1566} & \textbf{\scriptsize 0.2411} & {\scriptsize 0.0705} & {\scriptsize 0.1091}\tabularnewline
 & {\scriptsize mid-age} & \textbf{\scriptsize 0.3} & {\scriptsize 0.0226} & {\scriptsize 0.042} & \textbf{\scriptsize 0.1441} & {\scriptsize 0.1278} & {\scriptsize 0.1355} & \textbf{\scriptsize 0.135} & {\scriptsize 0.1654} & {\scriptsize 0.1486} & \textbf{\scriptsize 0.1897} & {\scriptsize 0.0827} & {\scriptsize 0.1152}\tabularnewline
 & {\scriptsize elder} & \textbf{\scriptsize 0.5} & {\scriptsize 0.0167} & {\scriptsize 0.0323} & \textbf{\scriptsize 0.0857} & {\scriptsize 0.1} & {\scriptsize 0.0923} & \textbf{\scriptsize 0.0792} & {\scriptsize 0.1333} & {\scriptsize 0.0994} & \textbf{\scriptsize 0.0476} & {\scriptsize 0.0167} & {\scriptsize 0.0247}\tabularnewline
\cline{2-14} 
 & {\scriptsize Macro-avg.} & \textbf{\scriptsize 0.4275} & {\scriptsize 0.2211} & {\scriptsize 0.2915} & \textbf{\scriptsize 0.3047} & {\scriptsize 0.1497} & {\scriptsize 0.2008} & \textbf{\scriptsize 0.3075} & {\scriptsize 0.1569} & {\scriptsize 0.2078} & \textbf{\scriptsize 0.3247} & {\scriptsize 0.1001} & {\scriptsize 0.153}\tabularnewline
 & {\scriptsize Micro-avg.} & \textbf{\scriptsize 0.4486} & {\scriptsize 0.4195} & {\scriptsize 0.4336} & \textbf{\scriptsize 0.3932} & {\scriptsize 0.1802} & {\scriptsize 0.2472} & \textbf{\scriptsize 0.3743} & {\scriptsize 0.1715} & {\scriptsize 0.2353} & \textbf{\scriptsize 0.4623} & {\scriptsize 0.1404} & {\scriptsize 0.2154}\tabularnewline
\hline 
{\scriptsize \multirow{7}{*}{religious affiliation}} & {\scriptsize atheist} & \textbf{\scriptsize 1} & {\scriptsize 0.2576} & {\scriptsize 0.4096} & \textbf{\scriptsize 0.4719} & {\scriptsize 0.6364} & {\scriptsize 0.5419} & \textbf{\scriptsize 0.4699} & {\scriptsize 0.5909} & {\scriptsize 0.5235} & \textbf{\scriptsize 0.6579} & {\scriptsize 0.3788} & {\scriptsize 0.4808}\tabularnewline
 & {\scriptsize budhist} & \textbf{\scriptsize 1} & {\scriptsize 0.2195} & {\scriptsize 0.36} & \textbf{\scriptsize 0.6667} & {\scriptsize 0.6341} & {\scriptsize 0.65} & \textbf{\scriptsize 0.4143} & {\scriptsize 0.7073} & {\scriptsize 0.5225} & \textbf{\scriptsize 0.5769} & {\scriptsize 0.3659} & {\scriptsize 0.4478}\tabularnewline
 & {\scriptsize christian} & {\scriptsize 0.9174} & {\scriptsize 0.9186} & {\scriptsize 0.918} & \textbf{\scriptsize 0.9899} & {\scriptsize 0.7896} & {\scriptsize 0.8785} & \textbf{\scriptsize 0.9926} & {\scriptsize 0.7409} & {\scriptsize 0.8485} & \textbf{\scriptsize 0.9928} & {\scriptsize 0.768} & {\scriptsize 0.8661}\tabularnewline
 & {\scriptsize jewish} & \textbf{\scriptsize 0.9592} & {\scriptsize 0.5109} & {\scriptsize 0.6667} & \textbf{\scriptsize 0.6495} & {\scriptsize 0.6848} & {\scriptsize 0.6667} & \textbf{\scriptsize 0.617} & {\scriptsize 0.6304} & {\scriptsize 0.6237} & \textbf{\scriptsize 0.8154} & {\scriptsize 0.5761} & {\scriptsize 0.6752}\tabularnewline
 & {\scriptsize muslim} & \textbf{\scriptsize 1} & {\scriptsize 0.4571} & {\scriptsize 0.6275} & \textbf{\scriptsize 0.8} & {\scriptsize 0.6857} & {\scriptsize 0.7385} & \textbf{\scriptsize 0.2747} & {\scriptsize 0.7143} & {\scriptsize 0.3968} & \textbf{\scriptsize 0.8571} & {\scriptsize 0.5143} & {\scriptsize 0.6429}\tabularnewline
\cline{2-14} 
 & {\scriptsize Macro-avg.} & \textbf{\scriptsize 0.9753} & {\scriptsize 0.4727} & {\scriptsize 0.6368} & \textbf{\scriptsize 0.7156} & {\scriptsize 0.6861} & {\scriptsize 0.7005} & \textbf{\scriptsize 0.5537} & {\scriptsize 0.6768} & {\scriptsize 0.6091} & \textbf{\scriptsize 0.78} & {\scriptsize 0.5206} & {\scriptsize 0.6245}\tabularnewline
 & {\scriptsize Micro-avg.} & \textbf{\scriptsize 0.9207} & {\scriptsize 0.8507} & {\scriptsize 0.8843} & \textbf{\scriptsize 0.927} & {\scriptsize 0.7736} & {\scriptsize 0.8434} & \textbf{\scriptsize 0.8734} & {\scriptsize 0.7288} & {\scriptsize 0.7946} & \textbf{\scriptsize 0.9658} & {\scriptsize 0.731} & {\scriptsize 0.8322}\tabularnewline
\hline 
{\scriptsize \multirow{4}{*}{political orientation}} & {\scriptsize democrat} & \textbf{\scriptsize 1} & {\scriptsize 0.26} & {\scriptsize 0.4127} & \textbf{\scriptsize 0.85} & {\scriptsize 0.34} & {\scriptsize 0.4857} & \textbf{\scriptsize 0.6905} & {\scriptsize 0.58} & {\scriptsize 0.6304} & \textbf{\scriptsize 0.7045} & {\scriptsize 0.62} & {\scriptsize 0.6596}\tabularnewline
 & {\scriptsize republican} & {\scriptsize 0.9157} & {\scriptsize 0.9583} & {\scriptsize 0.9365} & {\scriptsize 0.944} & {\scriptsize 0.9093} & {\scriptsize 0.9263} & {\scriptsize 0.973} & {\scriptsize 0.8848} & {\scriptsize 0.9268} & \textbf{\scriptsize 0.9808} & {\scriptsize 0.875} & {\scriptsize 0.9249}\tabularnewline
\cline{2-14} 
 & {\scriptsize Macro-avg.} & \textbf{\scriptsize 0.9579} & {\scriptsize 0.6092} & {\scriptsize 0.7447} & \textbf{\scriptsize 0.897} & {\scriptsize 0.6247} & {\scriptsize 0.7365} & \textbf{\scriptsize 0.8318} & {\scriptsize 0.7324} & {\scriptsize 0.7789} & \textbf{\scriptsize 0.8427} & {\scriptsize 0.7475} & {\scriptsize 0.7922}\tabularnewline
 & {\scriptsize Micro-avg.} & \textbf{\scriptsize 0.9182} & {\scriptsize 0.8821} & {\scriptsize 0.8998} & \textbf{\scriptsize 0.9395} & {\scriptsize 0.8472} & {\scriptsize 0.8909} & \textbf{\scriptsize 0.9443} & {\scriptsize 0.8515} & {\scriptsize 0.8955} & \textbf{\scriptsize 0.951} & {\scriptsize 0.8472} & {\scriptsize 0.8961}\tabularnewline
\hline 
{\scriptsize \multirow{4}{*}{sexual orientation}} & {\scriptsize heterosexual} & {\scriptsize 1} & {\scriptsize 0} & {\scriptsize 0} & {\scriptsize 1} & {\scriptsize 0} & {\scriptsize 0} & {\scriptsize 1} & {\scriptsize 0} & {\scriptsize 0} & {\scriptsize 1} & {\scriptsize 0} & {\scriptsize 0}\tabularnewline
 & {\scriptsize homosexual} & \emph{\scriptsize 0.9892} & {\scriptsize 0.9418} & {\scriptsize 0.9649} & {\scriptsize 1} & {\scriptsize 0.8116} & {\scriptsize 0.896} & {\scriptsize 1} & {\scriptsize 0.8116} & {\scriptsize 0.896} & {\scriptsize 1} & {\scriptsize 0.8116} & {\scriptsize 0.896}\tabularnewline
\cline{2-14} 
 & {\scriptsize Macro-avg.} & \textbf{\scriptsize 0.9946} & {\scriptsize 0.4709} & {\scriptsize 0.6392} & \textbf{\scriptsize 1} & {\scriptsize 0.4058} & {\scriptsize 0.5773} & \textbf{\scriptsize 1} & {\scriptsize 0.4058} & {\scriptsize 0.5773} & \textbf{\scriptsize 1} & {\scriptsize 0.4058} & {\scriptsize 0.5773}\tabularnewline
 & {\scriptsize Micro-avg.} & {\scriptsize 0.9892} & {\scriptsize 0.9322} & {\scriptsize 0.9599} & {\scriptsize 1} & {\scriptsize 0.8034} & {\scriptsize 0.891} & {\scriptsize 1} & {\scriptsize 0.8034} & {\scriptsize 0.891} & {\scriptsize 1} & {\scriptsize 0.8034} & {\scriptsize 0.891}\tabularnewline
\hline 
{\scriptsize \multirow{8}{*}{race/ethnicity}} & {\scriptsize asian-american} & \textbf{\scriptsize 0.8571} & {\scriptsize 0.4615} & {\scriptsize 0.6} & \textbf{\scriptsize 0.8571} & {\scriptsize 0.4615} & {\scriptsize 0.6} & \textbf{\scriptsize 0.75} & {\scriptsize 0.4615} & {\scriptsize 0.5714} & \textbf{\scriptsize 0.8571} & {\scriptsize 0.4615} & {\scriptsize 0.6}\tabularnewline
 & {\scriptsize black} & \textbf{\scriptsize 0.9412} & {\scriptsize 0.7805} & {\scriptsize 0.8533} & \textbf{\scriptsize 0.9412} & {\scriptsize 0.7805} & {\scriptsize 0.8533} & \textbf{\scriptsize 0.9412} & {\scriptsize 0.7805} & {\scriptsize 0.8533} & \textbf{\scriptsize 0.9412} & {\scriptsize 0.7805} & {\scriptsize 0.8533}\tabularnewline
 & {\scriptsize hispanic} & {\scriptsize 1} & {\scriptsize 0} & {\scriptsize 0} & {\scriptsize 1} & {\scriptsize 0} & {\scriptsize 0} & {\scriptsize 1} & {\scriptsize 0} & {\scriptsize 0} & {\scriptsize 1} & {\scriptsize 0} & {\scriptsize 0}\tabularnewline
 & {\scriptsize native-american} & \textbf{\scriptsize 0.9231} & {\scriptsize 0.75} & {\scriptsize 0.8276} & \textbf{\scriptsize 0.9231} & {\scriptsize 0.75} & {\scriptsize 0.8276} & \textbf{\scriptsize 0.9167} & {\scriptsize 0.6875} & {\scriptsize 0.7857} & \textbf{\scriptsize 0.9231} & {\scriptsize 0.75} & {\scriptsize 0.8276}\tabularnewline
 & {\scriptsize native-hawaiian} & {\scriptsize 1} & {\scriptsize 0} & {\scriptsize 0} & {\scriptsize 1} & {\scriptsize 0} & {\scriptsize 0} & {\scriptsize 1} & {\scriptsize 0} & {\scriptsize 0} & {\scriptsize 1} & {\scriptsize 0} & {\scriptsize 0}\tabularnewline
 & {\scriptsize white} & {\scriptsize 1} & {\scriptsize 0} & {\scriptsize 0} & {\scriptsize 1} & {\scriptsize 0} & {\scriptsize 0} & {\scriptsize 1} & {\scriptsize 0} & {\scriptsize 0} & {\scriptsize 1} & {\scriptsize 0} & {\scriptsize 0}\tabularnewline
\cline{2-14} 
 & {\scriptsize Macro-avg.} & \textbf{\scriptsize 0.9536} & {\scriptsize 0.332} & {\scriptsize 0.4925} & \textbf{\scriptsize 0.9536} & {\scriptsize 0.332} & {\scriptsize 0.4925} & \textbf{\scriptsize 0.9347} & {\scriptsize 0.3216} & {\scriptsize 0.4785} & \textbf{\scriptsize 0.9536} & {\scriptsize 0.332} & {\scriptsize 0.4925}\tabularnewline
 & {\scriptsize Micro-avg.} & \textbf{\scriptsize 0.9259} & {\scriptsize 0.641} & {\scriptsize 0.7576} & \textbf{\scriptsize 0.9259} & {\scriptsize 0.641} & {\scriptsize 0.7576} & \textbf{\scriptsize 0.9074} & {\scriptsize 0.6282} & {\scriptsize 0.7424} & \textbf{\scriptsize 0.9259} & {\scriptsize 0.641} & {\scriptsize 0.7576}\tabularnewline
\hline
\end{tabular}
\end{sidewaystable*}

\par\end{center}{\scriptsize \par}

\noindent \begin{center}
{\scriptsize }%
\begin{sidewaystable*}
{\scriptsize \caption{Performance figures of the four {}``flavors'' of the \emph{McC-Splat}
algorithm working on the \emph{WeFollow} dataset. Bold figures correspond
to performance differences above 10\% when comparing against the random
classifier (see table \ref{RandomClassifierWeFollow}).}
\label{TMcCSplatWeFollow}}{\scriptsize \par}

{\scriptsize }\begin{tabular}{|c|c|ccc|ccc|ccc|ccc|}
\hline 
{\scriptsize \multirow{2}{*}{}\textbf{\scriptsize Attribute}{\scriptsize }} & {\scriptsize \multirow{2}{*}{}\textbf{\scriptsize Class}{\scriptsize }} & \multicolumn{3}{c|}{\textbf{\scriptsize Plain-vanilla}} & \multicolumn{3}{c|}{\textbf{\scriptsize Sink-absolute}} & \multicolumn{3}{c|}{\textbf{\scriptsize Sink-relative}} & \multicolumn{3}{c|}{\textbf{\scriptsize Percentile}}\tabularnewline
\cline{3-14} 
 &  & {\scriptsize P} & {\scriptsize R} & {\scriptsize $F_{1}$} & {\scriptsize P} & {\scriptsize R} & {\scriptsize $F_{1}$} & {\scriptsize P} & {\scriptsize R} & {\scriptsize $F_{1}$} & {\scriptsize P} & {\scriptsize R} & {\scriptsize $F_{1}$}\tabularnewline
\hline
\hline 
{\scriptsize \multirow{7}{*}{religious affiliation}} & {\scriptsize atheist} & \textbf{\scriptsize 0.9795} & {\scriptsize 0.1496} & {\scriptsize 0.2595} & \textbf{\scriptsize 0.8615} & {\scriptsize 0.3117} & {\scriptsize 0.4577} & \textbf{\scriptsize 0.8355} & {\scriptsize 0.3062} & {\scriptsize 0.4481} & \textbf{\scriptsize 0.8735} & {\scriptsize 0.2326} & {\scriptsize 0.3673}\tabularnewline
 & {\scriptsize budhist} & \textbf{\scriptsize 0.95} & {\scriptsize 0.0759} & {\scriptsize 0.1406} & \textbf{\scriptsize 0.7349} & {\scriptsize 0.1625} & {\scriptsize 0.2661} & \textbf{\scriptsize 0.5447} & {\scriptsize 0.1864} & {\scriptsize 0.2778} & \textbf{\scriptsize 0.7034} & {\scriptsize 0.1358} & {\scriptsize 0.2277}\tabularnewline
 & {\scriptsize christian} & \textbf{\scriptsize 0.8705} & {\scriptsize 0.3084} & {\scriptsize 0.4555} & \textbf{\scriptsize 0.9878} & {\scriptsize 0.2643} & {\scriptsize 0.417} & \textbf{\scriptsize 0.9982} & {\scriptsize 0.2411} & {\scriptsize 0.3897} & \textbf{\scriptsize 0.9983} & {\scriptsize 0.2518} & {\scriptsize 0.4021}\tabularnewline
 & {\scriptsize jewish} & \textbf{\scriptsize 0.9672} & {\scriptsize 0.1664} & {\scriptsize 0.284} & \textbf{\scriptsize 0.6524} & {\scriptsize 0.2144} & {\scriptsize 0.3227} & \textbf{\scriptsize 0.6329} & {\scriptsize 0.2116} & {\scriptsize 0.3171} & \textbf{\scriptsize 0.768} & {\scriptsize 0.1961} & {\scriptsize 0.3124}\tabularnewline
 & {\scriptsize muslim} & \textbf{\scriptsize 0.973} & {\scriptsize 0.0633} & {\scriptsize 0.1188} & \textbf{\scriptsize 0.6667} & {\scriptsize 0.0984} & {\scriptsize 0.1715} & \textbf{\scriptsize 0.2331} & {\scriptsize 0.109} & {\scriptsize 0.1485} & \textbf{\scriptsize 0.6849} & {\scriptsize 0.0879} & {\scriptsize 0.1558}\tabularnewline
\cline{2-14} 
 & {\scriptsize Macro-avg.} & \textbf{\scriptsize 0.948} & {\scriptsize 0.1527} & {\scriptsize 0.2631} & \textbf{\scriptsize 0.7807} & {\scriptsize 0.2103} & {\scriptsize 0.3313} & \textbf{\scriptsize 0.6489} & {\scriptsize 0.2111} & {\scriptsize 0.3185} & \textbf{\scriptsize 0.8056} & {\scriptsize 0.1808} & {\scriptsize 0.2954}\tabularnewline
 & {\scriptsize Micro-avg.} & \textbf{\scriptsize 0.8799} & {\scriptsize 0.2662} & {\scriptsize 0.4088} & \textbf{\scriptsize 0.9361} & {\scriptsize 0.2543} & {\scriptsize 0.4} & \textbf{\scriptsize 0.8768} & {\scriptsize 0.2382} & {\scriptsize 0.3746} & \textbf{\scriptsize 0.9566} & {\scriptsize 0.2351} & {\scriptsize 0.3774}\tabularnewline
\hline 
{\scriptsize \multirow{4}{*}{political orientation}} & {\scriptsize democrat} & \textbf{\scriptsize 1} & {\scriptsize 0.0478} & {\scriptsize 0.0913} & \textbf{\scriptsize 0.9429} & {\scriptsize 0.0686} & {\scriptsize 0.1279} & \textbf{\scriptsize 0.7899} & {\scriptsize 0.1954} & {\scriptsize 0.3133} & \textbf{\scriptsize 0.7638} & {\scriptsize 0.2017} & {\scriptsize 0.3191}\tabularnewline
 & {\scriptsize republican} & {\scriptsize 0.9178} & {\scriptsize 0.3717} & {\scriptsize 0.5291} & {\scriptsize 0.9338} & {\scriptsize 0.3602} & {\scriptsize 0.5199} & \textbf{\scriptsize 0.9778} & {\scriptsize 0.3536} & {\scriptsize 0.5194} & \textbf{\scriptsize 0.9831} & {\scriptsize 0.3502} & {\scriptsize 0.5164}\tabularnewline
\cline{2-14} 
 & {\scriptsize Macro-avg.} & \textbf{\scriptsize 0.9589} & {\scriptsize 0.2098} & {\scriptsize 0.3442} & \textbf{\scriptsize 0.9384} & {\scriptsize 0.2144} & {\scriptsize 0.349} & \textbf{\scriptsize 0.8839} & {\scriptsize 0.2745} & {\scriptsize 0.4189} & \textbf{\scriptsize 0.8735} & {\scriptsize 0.276} & {\scriptsize 0.4194}\tabularnewline
 & {\scriptsize Micro-avg.} & \textbf{\scriptsize 0.9191} & {\scriptsize 0.3324} & {\scriptsize 0.4882} & \textbf{\scriptsize 0.934} & {\scriptsize 0.3248} & {\scriptsize 0.482} & \textbf{\scriptsize 0.9616} & {\scriptsize 0.3344} & {\scriptsize 0.4963} & \textbf{\scriptsize 0.9627} & {\scriptsize 0.3322} & {\scriptsize 0.4939}\tabularnewline
\hline 
{\scriptsize \multirow{4}{*}{sexual orientation}} & {\scriptsize heterosexual} & {\scriptsize 1} & {\scriptsize 1} & {\scriptsize 1} & {\scriptsize 1} & {\scriptsize 1} & {\scriptsize 1} & {\scriptsize 0} & {\scriptsize 1} & {\scriptsize 0} & {\scriptsize 0} & {\scriptsize 1} & {\scriptsize 0}\tabularnewline
 & {\scriptsize homosexual} & {\scriptsize 1} & {\scriptsize 0.2685} & {\scriptsize 0.4234} & {\scriptsize 1} & {\scriptsize 0.2415} & {\scriptsize 0.389} & {\scriptsize 1} & {\scriptsize 0.2402} & {\scriptsize 0.3874} & {\scriptsize 1} & {\scriptsize 0.2402} & {\scriptsize 0.3874}\tabularnewline
\cline{2-14} 
 & {\scriptsize Macro-avg.} & {\scriptsize 1} & {\scriptsize 0.6343} & {\scriptsize 0.7762} & {\scriptsize 1} & {\scriptsize 0.6208} & {\scriptsize 0.766} & {\scriptsize 0.5} & {\scriptsize 0.6201} & {\scriptsize 0.5536} & {\scriptsize 0.5} & {\scriptsize 0.6201} & {\scriptsize 0.5536}\tabularnewline
 & {\scriptsize Micro-avg.} & {\scriptsize 1} & {\scriptsize 0.2685} & {\scriptsize 0.4234} & {\scriptsize 1} & {\scriptsize 0.2415} & {\scriptsize 0.389} & {\scriptsize 0.9947} & {\scriptsize 0.2402} & {\scriptsize 0.387} & {\scriptsize 0.9965} & {\scriptsize 0.2402} & {\scriptsize 0.3871}\tabularnewline
\hline 
{\scriptsize \multirow{8}{*}{race/ethnicity}} & {\scriptsize asian-american} & \textbf{\scriptsize 0.8571} & {\scriptsize 0.035} & {\scriptsize 0.0672} & \textbf{\scriptsize 0.8571} & {\scriptsize 0.035} & {\scriptsize 0.0672} & \textbf{\scriptsize 0.8571} & {\scriptsize 0.035} & {\scriptsize 0.0672} & \textbf{\scriptsize 0.8913} & {\scriptsize 0.0341} & {\scriptsize 0.0658}\tabularnewline
 & {\scriptsize black} & \textbf{\scriptsize 0.6818} & {\scriptsize 0.1624} & {\scriptsize 0.2623} & \textbf{\scriptsize 0.6818} & {\scriptsize 0.1624} & {\scriptsize 0.2623} & \textbf{\scriptsize 0.6919} & {\scriptsize 0.161} & {\scriptsize 0.2613} & \textbf{\scriptsize 0.6722} & {\scriptsize 0.1637} & {\scriptsize 0.2633}\tabularnewline
 & {\scriptsize hispanic} & {\scriptsize 1} & {\scriptsize 0} & {\scriptsize 0} & {\scriptsize 1} & {\scriptsize 0} & {\scriptsize 0} & {\scriptsize 1} & {\scriptsize 0} & {\scriptsize 0} & {\scriptsize 1} & {\scriptsize 0} & {\scriptsize 0}\tabularnewline
 & {\scriptsize native-american} & \textbf{\scriptsize 0.4828} & {\scriptsize 0.0897} & {\scriptsize 0.1514} & \textbf{\scriptsize 0.4828} & {\scriptsize 0.0897} & {\scriptsize 0.1514} & \textbf{\scriptsize 0.4375} & {\scriptsize 0.0897} & {\scriptsize 0.1489} & \textbf{\scriptsize 0.4483} & {\scriptsize 0.0833} & {\scriptsize 0.1405}\tabularnewline
 & {\scriptsize native-hawaiian} & {\scriptsize 0} & {\scriptsize 1} & {\scriptsize 0} & {\scriptsize 0} & {\scriptsize 1} & {\scriptsize 0} & {\scriptsize 0} & {\scriptsize 1} & {\scriptsize 0} & {\scriptsize 0} & {\scriptsize 1} & {\scriptsize 0}\tabularnewline
 & {\scriptsize white} & {\scriptsize 0} & {\scriptsize 1} & {\scriptsize 0} & {\scriptsize 0} & {\scriptsize 1} & {\scriptsize 0} & {\scriptsize 0} & {\scriptsize 1} & {\scriptsize 0} & {\scriptsize 0} & {\scriptsize 1} & {\scriptsize 0}\tabularnewline
\cline{2-14} 
 & {\scriptsize Macro-avg.} & \textbf{\scriptsize 0.5036} & {\scriptsize 0.3812} & {\scriptsize 0.4339} & \textbf{\scriptsize 0.5036} & {\scriptsize 0.3812} & {\scriptsize 0.4339} & \textbf{\scriptsize 0.4978} & {\scriptsize 0.381} & {\scriptsize 0.4316} & \textbf{\scriptsize 0.502} & {\scriptsize 0.3802} & {\scriptsize 0.4327}\tabularnewline
 & {\scriptsize Micro-avg.} & \textbf{\scriptsize 0.7078} & {\scriptsize 0.0547} & {\scriptsize 0.1015} & \textbf{\scriptsize 0.7078} & {\scriptsize 0.0547} & {\scriptsize 0.1015} & \textbf{\scriptsize 0.7045} & {\scriptsize 0.0544} & {\scriptsize 0.101} & \textbf{\scriptsize 0.7036} & {\scriptsize 0.0541} & {\scriptsize 0.1006}\tabularnewline
\hline
\end{tabular}
\end{sidewaystable*}

\par\end{center}
\end{document}